\documentclass[final,3p,twocolumn]{elsarticle}
\usepackage{graphicx}

\usepackage{amsmath}
\usepackage{amssymb}
\usepackage{amsbsy}
\usepackage{epstopdf}
\usepackage{bm}

\journal{Computers \& Fluids}

\begin{document}

\begin{frontmatter}
\title{The consequences of finite-time proper orthogonal decomposition for an extensively chaotic flow field}

\author[tamu]{Andrew Duggleby\corref{cor1}}
\ead{aduggleby@tamu.edu}
\author[vt]{Mark R. Paul}

\cortext[cor1]{Corresponding author}
\address[tamu]{Department of Mechanical Engineering, 
        Texas A\&M University,
        College Station, TX  77843-3123}  
 \address[vt]{Department of Mechanical Engineering, Virginia Polytechinic Institute \& State University, Blacksburg, VA 24061}
\begin{abstract}
The use of proper orthogonal decomposition (POD) to explore the complex 
fluid flows that are common in engineering applications is increasing and has yielded new physical 
insights.  However, for most engineering systems the dimension of the dynamics is expected to 
be very large yet the flow field data is available only for a finite time.  In this context, it is important 
to establish the convergence of the POD in order to accurately estimate such quantities as the 
Karhunen-Lo\`{e}ve dimension. Using direct numerical simulations of Rayleigh-B\'{e}nard 
convection in a finite cylindrical geometry we explore a regime exhibiting extensive chaos and 
demonstrate the consequences of performing a POD with a finite amount of data. In 
particular, we show that the convergence in time of the eigenvalue spectrum, the eigenfunctions, 
and the dimension are very slow in comparison with the time scale of the convection rolls and that 
the errors incurred by not using the asymptotic values can be significant.   We compute the 
dimension using two approaches, the method of snapshots and a Fourier method 
that exploits the azimuthal symmetry. We find that the convergence rate of the Fourier method 
is vastly improved over the method of snapshots. The dimension is found to be extensive 
as the system size is increased and for a dimension measurement that captures 90\% of 
the variance in the data the Karhunen-Lo\`{e}ve dimension is about 20 times larger 
than the Lyapunov dimension.
\end{abstract}

\end{frontmatter}

\section{Introduction}
Many open challenges remain in the development of ways to describe, model, and 
predict chaotic fluid flows~\cite{cross:1993}.  One method proposed by Lumley~\cite{lumley1} 
to study turbulence is to use a Karhunen-Lo\`{e}ve decomposition, or proper orthogonal 
decomposition (POD), to decompose the flow into an optimal set of basis functions.  
This has lead to many discoveries and has increased our understanding of  
turbulence including the dynamics of streaks and bursting events~\cite{sirovichKL1,sirovichKL2,sirovichKL3,sirovich_chaos,sirovich1,sirovich2,ball,duggleby_JoT,Duggleby2008PhilTrans}, the dynamics of energy transfer~\cite{webber2,zhou_sirovich}, and the mechanism of drag 
reduction~\cite{housiadas,duggleby_drPipe}.  Recently, this has been extended 
to many experimental and numerical investigations of more complicated flow fields in engineering 
applications that are quite different than a typical theoretical investigation of fluid 
turbulence (c.f.~\cite{lian:2003,epureanu:2001}). An important difference is that the 
interval of time over which the data is available is often very short.  This limiting factor 
is present in both experimental and numerical studies.  It is anticipated that the dimension 
describing the dynamics of these systems is very large and as a result the short time 
observations provide only a limited sample of the overall dynamics.

We explore this by quantifying the convergence of the Karhunen-Lo\`{e}ve 
dimension with time for a high-dimensional fluid flow field.  Our results provide insights for 
the use of POD for more complex engineering flow fields where only a finite 
amount of data is available.  Our results show that (i) the eigenvalue spectrum, 
the eigenfunctions, and the dimension for the POD of a complex flow field can 
be approximated from finite time observations; and~(ii) tailoring the numerical 
algorithm to exploit the rotational invariance in the problem vastly improves the 
rate of convergence.

\begin{figure}[htb]
  \begin{center}
    \includegraphics[height=2.5in]{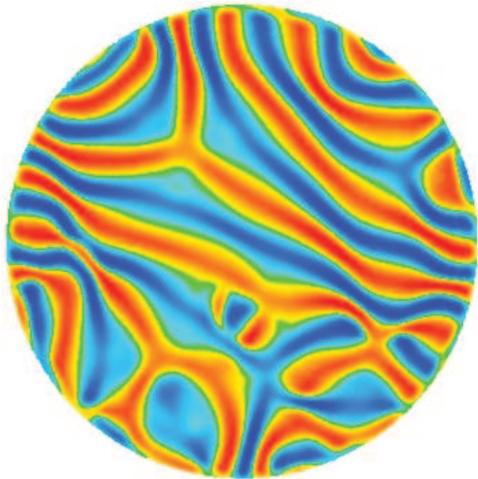}
  \end{center}
\caption{(color) A typical pattern of convection rolls from a numerical simulation of 
chaotic Rayleigh-B\'{e}nard convection in a shallow cylindrical domain. The flow field is shown 
at mid-depth $z=1/2$ where red indicates warm rising fluid and blue indicates cool 
falling fluid. For this simulation $\Gamma=10$,  $R/R_c=3.5$, and $\sigma=1$. }
\label{rbc}
\end{figure}

We show this by numerically integrating the time-dependent and three-dimensional 
Boussinesq equations that govern the fluid motion of Rayleigh-B\'{e}nard convection in a 
shallow cylindrical domain.  In particular, we explore the spatiotemporal chaos of the 
fluid convection rolls that arise when a layer of fluid is heated uniformly from below 
in a gravitational field.  A typical flow field pattern from our numerical simulations is 
shown in Fig.~\ref{rbc} which illustrates a horizontal mid-plane slice of the domain where 
red is hot rising fluid and blue is cool falling fluid.

The dimension of the attractor describing the dynamics of chaotic convection is 
expected to be much lower than one would find in fully developed turbulent 
flow~\cite{cross:1993,egolf:2000,paul2007}. However, the dimension of convection 
is large enough to test these ideas~\cite{paul2007}. Our results provide quantitative 
estimates for the convergence rates and magnitudes of important diagnostics that 
result from a POD study.

\section{Numerical Methods}
\subsection{Direct Numerical Simulation}
The fluid motion of Rayleigh-B\'{e}nard convection is governed by the Boussinesq equations
\begin{eqnarray}
\label{NS}
\hspace{-2em} \mathrm{\sigma}^{-1} \left( \partial_t + \mathbf{u} \cdot \mathbf{\nabla}
\right) \mathbf{u}\! &=& \! -\nabla P + \nabla^2 \mathbf{u} + \mathrm{R} T \hat{z}, \\
\label{energy}
\left( \partial_t + \mathbf{u} \cdot \mathbf{\nabla}
\right) T& = &\nabla^2 T, \\ 
\label{cont}
\mathbf{\nabla} \cdot \mathbf{u}&=&0,
\end{eqnarray}
which represent the conservation of momentum, energy, and mass. In our notation $\partial_t$ is 
a time derivative, $\hat{z}$ is a unit vector in the direction opposing gravity, $\mathbf{u}$ 
is the fluid velocity, $p$ is the pressure, and $T$ is the temperature.  The equations are 
nondimensionalized using the layer depth $d$ as the length scale, the vertical diffusion 
time of heat $\tau_v = d^2/\alpha$ for the time scale where $\alpha$ is the thermal diffusivity, 
and the constant temperature difference $\Delta T$ between the bottom and top plates 
as the temperature scale. In our investigation we consider a shallow cylindrical convection 
domain of radius $r_0$ and depth $d$. On all material walls we impose the no-slip boundary 
condition 
\begin{equation}
\mathbf{u}=0.
\end{equation}
The top and bottom plates are held at a constant temperature $T(z=0)=1$ and $T(z=1)=0$ and 
the lateral sidewalls of the domain are perfectly conducting boundaries with 
\begin{equation}
 T = 1 - z.
\end{equation}
There are three important nondimensional parameters that completely describe the 
dynamics.  The aspect ratio of the domain
\begin{equation}
\Gamma = \frac{r_0}{d}
\end{equation}
which is a measure of the system size and $r_0$ is the radius of the domain. The Prandtl number
\begin{equation}
\sigma= \frac{\nu}{\alpha}
\end{equation} 
where $\nu$ is the kinematic viscosity, and the Rayleigh number
\begin{equation}
R = \frac{g \beta \Delta T d^3}{\alpha \nu}
\end{equation}
where $g$ is the acceleration due to gravity and $\beta$ is the coefficient of thermal expansion.

The control parameter that is varied in most experiments is $R$~\cite{bodenschatz:2000}. For 
no-slip boundaries the critical value of the Rayleigh number is $R_c=1708$ which corresponds to 
the onset of convection rolls. For a fluid with $\sigma \approx 1$, as $R$ is increased 
$R \lesssim 10^4$ the steady convection rolls are 
replaced by patterns of rolls with time dependent dynamics that include periodic, quasiperiodic,
and chaotic dynamics~(c.f.~\cite{ahlers:1974,libchaber:1978,pocheau:1988,paul:2001}). As $R$ 
is increased further $R \gtrsim 10^5$ the convection rolls are annihilated and replaced by thermal 
plume structures yielding turbulent convection~\cite{cross:1993}. 
In this paper we are interested in the range where the patterns of convection rolls exhibit 
spatiotemporal chaos and we use $R=6000$ (which yields $R/R_c=3.5$), $\sigma=1$ and 
a range of system sizes $6 \le \Gamma \le 15$.  In the 
following we solve the time-dependent Boussinesq equations using a geometrically flexible, efficient, 
spectral element algorithm~\cite{fischer:1994,fischer:1997,tufo}.  Further details on 
the specific application of this numerical method to thermal convection can be found in 
Ref.~\cite{paul:2002:physd}.

\subsection{Proper Orthogonal Decomposition}
The POD of a fluid flow field can be cast as the solution of the Fredholm integral,
\begin{equation}
\label{KL}
\int_\Omega K(\mathbf{x}, \mathbf{x}') \bm{\Phi}
(\mathbf{x}') d^3x' = \mu \bm{\Phi}(\mathbf{x}),
\end{equation}
where,
\begin{equation}
\label{two_point}
K(\mathbf{x}, \mathbf{x'})=\lim_{\tau\to\infty} \frac{1}{\tau} \int_0^\tau \mathbf{u} (\mathbf{x},t) \otimes \mathbf{u}
(\mathbf{x}',t) dt,
\end{equation}
and $\otimes$ is an outer product, $\Omega$ is the volume of the entire domain, 
$\mathbf{x}$ is the position vector, $\bm{\Phi}(\mathbf{x})$
is the eigenvector with associated eigenvalue $\mu$, and
$K(\mathbf{x}, \mathbf{x}')$ is the kernel.  The kernel is built using
the two-point correlation of the fluctuating space-time velocity components 
$\mathbf{u}(\mathbf{x}, t)$ averaged over time $\tau$. The Karhunen-Lo\`{e}ve 
dimension $D_{KL}$ is the number of eigenmodes necessary to capture a given 
fraction $f$ of the total variance of the data 
\begin{equation}
\frac{\displaystyle\sum_{j=1}^{D_{KL}} \mu_j}{\displaystyle\sum_{j=1}^\infty \mu_j} = f
\end{equation}
where a typical choice is $f=0.9$. The observation time $\tau$ should be long enough 
such that most of the dynamics on the attractor have been observed. In the 
limit $\tau \rightarrow \infty$ this is satisfied. However, for a finite amount of data 
it is desired that the data is sufficient for the dimension to have converged to a value 
close to its asymptotic value such that $D_{KL} \approx D_{\infty}$.

For the Rayleigh-B\'{e}nard convection studied here we perform a POD on the 
two-dimensional flow field pattern at the mid plane as shown in Fig.~\ref{rbc}. 
The reason for this is two-fold. One, a typical experiment using 
shadowgraphy to measure the convective roll pattern would generate the same 
type of data. Second, it is expected that the convection rolls do not contain structure in the vertical 
direction that would have a significant impact upon the dimension.  In collecting 
data for our POD we use images from the flow field separated in time 
by 5$\tau_v$ where the total duration of the numerical simulation depends upon 
the system size explored.

\subsubsection{Fourier Method}  
The size of the eigenvalue problem to solve the Fredholm integral in Eq.~(\ref{KL}) is large.  
Two methods that are used to reduce the computations are the Fourier method and the 
method of snapshots.

For a rotationally invariant system every rotation of the solution is also valid.  The 
coefficients of the POD in the azimuthal direction are therefore those of a 
Fourier series~\cite{Holmes}.  When this invariance is built into the computational procedure 
we refer to this as the Fourier method.  Therefore, the eigenmode 
for mode $m$,
\begin{equation}
\bm{\Phi}_m (r, \theta) = \bm{\Psi}_n^{q}(r)e^{in \theta}
\end{equation}
is uniquely described with an azimuthal wavenumber $n$ and eigen number $q$.  This 
reduces the computation of the Karhunen-Lo\`{e}ve modes to
\begin{equation}
\label{reduced_inner}
\int_0^{r_0} \mathcal{K}_n(r,r')\bm{\Psi}_n (r') r'
dr'=\mu_{n}\bm{\Psi}_n (r),
\end{equation}
where,
\begin{equation}
\label{reduced}
\mathcal{K}_n (r,r') = \lim_{\tau\to\infty} \frac{1}{\tau} \int_0^\tau \hat{\mathbf{u}}_n(r,t) \otimes
\hat{\mathbf{u}}_n^\star (r',t) dt,
\end{equation}
with $\star$ denoting the complex conjugate and the weight $r'$ is present 
because the inner product is evaluated in polar-cylindrical coordinates $(r,\theta)$.  The 
Fourier transform of the velocity in the azimuthal direction is $\hat{\mathbf{u}}_n(r,t)$. The 
kernel in Eq.~(\ref{reduced_inner}) is Hermitian as it was in 
Eq.~(\ref{KL}) and yields real and positive eigenvalues. Using a $Q$-point quadrature 
to solve Eq.~(\ref{reduced_inner}) yields $3Q$ eigenvectors and corresponding eigenvalues 
in descending order for each azimuthal mode $n$, denoted with eigen number $q$.  Physically, each eigenvector 
$\bm{\Psi}_n^q (r)e^{in \theta}$ represents a velocity field and the eigenvalue $\mu_n^q$ 
is the time averaged energy of that flow field.
 
\subsubsection{Method of Snapshots}
The method of snapshots recasts the eigenfunction in the Fredholm integral in Eq.~(\ref{KL}) as 
a linear combination of the snapshots where
\begin{equation}
\label{snapshots}
c(t)=\int_\Omega \mathbf{u}(\mathbf{x}',t) \bm{\Phi}(\mathbf{x}')d^3x'
\end{equation}
and Eq.~(\ref{KL}) is then
\begin{equation}
\lim_{\tau\to\infty} \frac{1}{\tau} \int_0^\tau \mathbf{u}(\mathbf{x},t) c(t) dt = \lambda \bm{\Phi}(\mathbf{x}).
\end{equation}
Multiplying both sides by $\mathbf{u}(\mathbf{x}, t')$ and integrating yields,  
\begin{equation}
\lim_{\tau\to\infty} \frac{1}{\tau} \int_0^\tau c(t) \int_\Omega \mathbf{u}(\mathbf{x}, t') \mathbf{u}(\mathbf{x},t) d^3x dt = \lambda c(t')
\end{equation}
or,
\begin{equation}
\label{snapshots_eq}
\lim_{\tau\to\infty} \frac{1}{\tau} \int_0^\tau  \left(\mathbf{u}(\mathbf{x}, t'), \mathbf{u}(\mathbf{x},t) \right) c(t) dt = \lambda c(t')
\end{equation}
where $(~,~)$ denotes an inner product over the domain $\Omega$.  The resulting equation is 
now a Fredholm integral in time. In fluid dynamics this is more computationally tractable because 
there are typically fewer time observations (snapshots) than there are spatial observations (grid points).
\begin{figure}[t!]
  \begin{center}
  \begin{tabular}{ll}
    \includegraphics[height=1.2in]{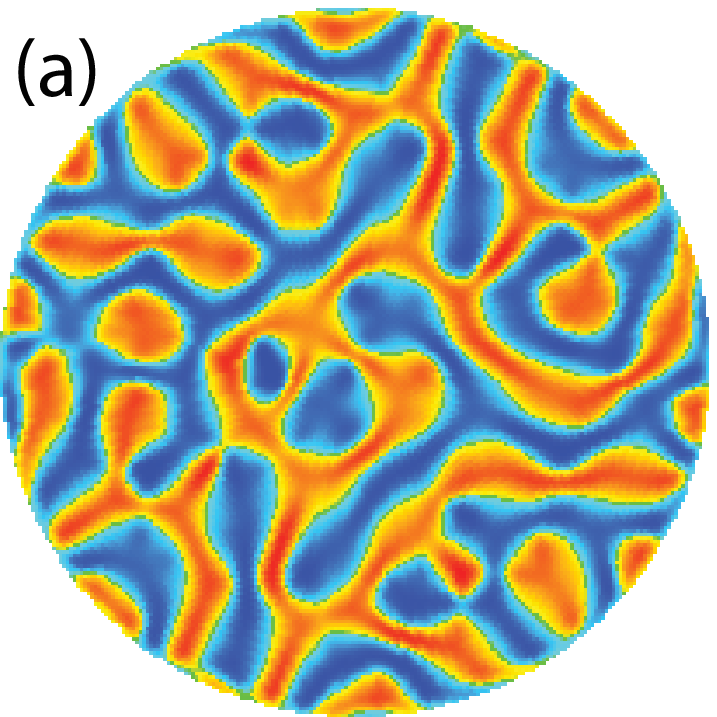} & \includegraphics[height=1.2in]{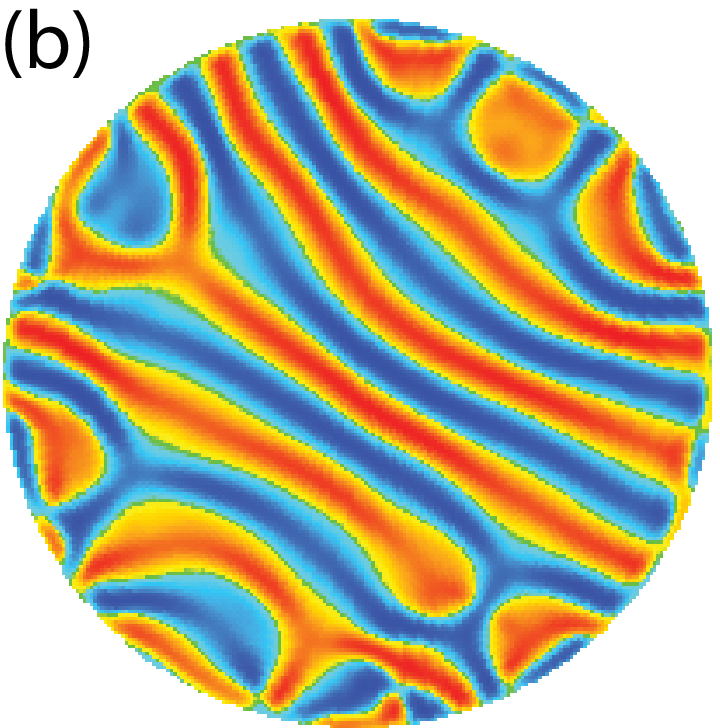} \\
    \includegraphics[height=1.2in]{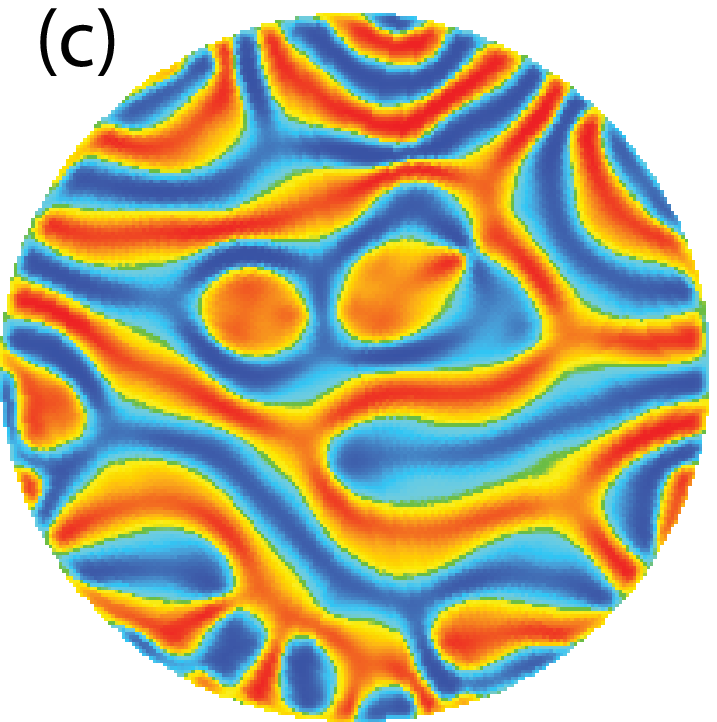} & \includegraphics[height=1.2in]{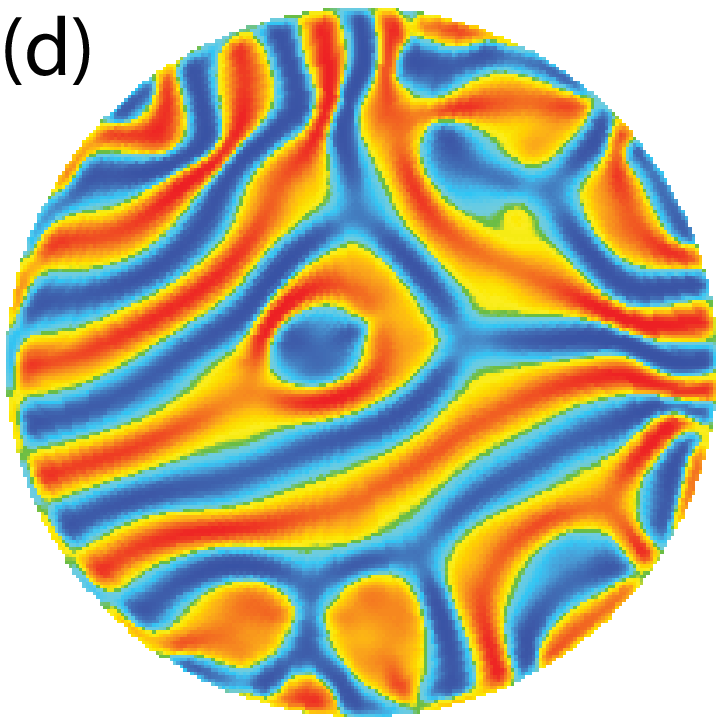} \\
    \end{tabular}
  \end{center}
\caption{(color) Four flow fields from numerical simulation that are separated in time 
by $50\tau_v$ for $\Gamma=10$, $R/R_c=3.5$, $\sigma=1$.  The values of $D_{KL}$ 
computed using the Fourier method for each individual flow field are 46.1, 46.8, 49.8, and 
48.5 for panels (a)-(d) respectively.}
\label{RBC_smalltest}
  \end{figure}
\begin{table}[t!]
\caption{The cumulative Karhunen-Lo\`{e}ve dimension of the four flow field images 
shown  in Fig.~\ref{RBC_smalltest}. The number of images used in computing the dimension 
is $n$ for both the method of snapshots and the Fourier method.}
\begin{center}
\begin{tabular}{lll}
\hline
Samples  & Snapshots & Fourier \\
  $n$ & $D_{KL}$ & $D_{KL}$ \\ 
\hline
  1 & 0.9 & 46.1 \\
  2 & 1.79 & 84.2 \\
  3 & 2.64 & 118.3 \\
  4 & 3.49 & 147.4 \\
  \hline
  \end{tabular}
  \end{center}
  \label{smalltest_dim}
\end{table}
\begin{figure}[t!]
\begin{center}
\includegraphics[height=2.5 in]{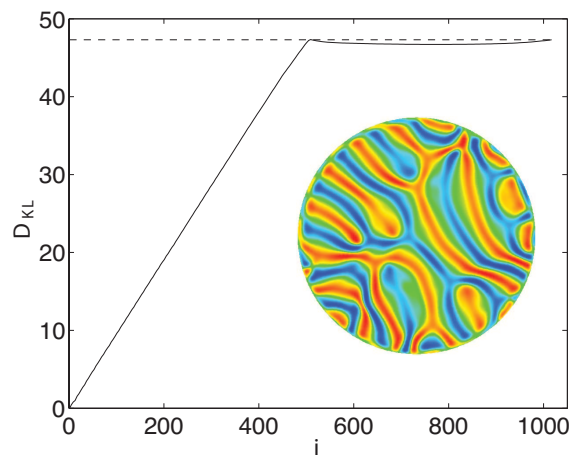}
\end{center}
\caption{(color)~The cumulative dimension of a flow field pattern where a single snapshot is 
rotated azimuthally by increments of $\Delta \theta = 2\pi /512$ to form new snapshots where 
the angle of rotation is $\theta_i = i \Delta \theta$ and $i$ is an integer. Results are shown for a 
total of 1024 rotations where method of snapshots is the solid line and the Fourier 
method is the dashed line.  The two methods agree when $i=512$ and $1024$ corresponding 
to one and two complete rotations of the flow field, respectively. The flow 
field used is at time $t=500$ (see inset) and for the simulation parameters 
$\Gamma=10$, $\sigma=1$, and $R/R_c=3.5$.}
\label{spintest}
\end{figure}

\section{Discussion}

\subsection{Comparison of the method of snapshots and the Fourier method}
We now illustrate the differences between the two methods through two examples using numerical data from 
our simulations for $\Gamma=10$. The flow field patterns from four instances in 
time separated by $50\tau_v$ are shown in Fig.~\ref{RBC_smalltest}.  Using the Fourier method the 
dimensions of these individual images are $D_{KL} = 46.1, 46.8, 49.8, 48.5$ for panels~(a)-(d), 
respectively.  We next construct a time series out of these images and compute the dimension using 
images from panels (a), (a,b), (a,b,c), and (a,b,c,d) which we refer to as the 
cumulative dimension in~Table \ref{smalltest_dim}.  The magnitude of the dimension increases rapidly with 
the addition of each new flow field image.  When using all four images the Fourier method yields a dimension 
of $D_{KL}=147.4$  which is less than the sum of the individual contributions 191.2.

The method of snapshots, which does not take advantage of the rotational invariance of the flow 
field, has only a dimension of 3.49 when using all four images. This comparison illustrates the rapid 
convergence of the Fourier method for the computation of $D_{KL}$ in high-dimensional systems.

Using the same convection domain with $\Gamma=10$ we now consider a single flow field image 
from time $t=500$ that is rotated azimuthally in small increments to create a series 
of images that is then used to compute $D_{KL}$. The specific flow field pattern that is used is 
shown in the inset of Fig.~\ref{spintest}. The image is rotated incrementally by 
$\Delta \theta = 2 \pi /512$ to yield rotations of $\theta = i \Delta \theta$ where $i=1,2,\ldots,1024$ 
resulting in two complete rotations.

The values of $D_{KL}$ from the method of snapshots 
and the Fourier method are shown in Fig.~\ref{spintest}. The dimension from the Fourier 
method is $D_{KL}=47.3$ using only one image and remains at this value for the subsequent 
incremental rotations.  The method of 
snapshots however, increases monotonically for each incremental rotation and agrees with the 
dimension from the Fourier method only after one complete rotation.  This occurs for 
subsequent complete rotations as well.

This example illustrates the clear computational advantage, in 
terms of rate of convergence,  of the Fourier method for computing $D_{KL}$. 
It also shows that both the method of snapshots and the Fourier method eventually yield the same 
value for $D_{KL}$.  The computed values of $D_{KL}$ only agree after the flow field pattern 
has undergone a complete rotation.  For the case of actual flow field data it is expected 
that the two approaches would also agree if enough data were used for the method 
of snapshots to sample from all of the possible orientations. However, this can be a very 
slow process as we will demonstrate. 

\subsection{The Lyapunov and Karhunen-Lo\`{e}ve dimensions}
A significant advantage of studying chaotic Rayleigh-B\'{e}nard convection is that the spectrum 
of Lyapunov exponents have been calculated to yield quantitative measurements of the 
Lyapunov dimension~\cite{egolf:2000,paul2007}.  The spectrum of Lyapunov exponents 
$\lambda_j$ measure the exponential separation of trajectories in phase space and are 
ordered from largest to smallest. A single positive exponent is the defining feature of 
deterministic chaos~(c.f. \cite{cross:1993,eckmann1985,wolf:1985}). The sum of the first $N$ 
exponents indicates the exponential growth of an $N-$dimensional ball of initial 
conditions in phase space. The number of exponents that must be added in order for their 
sum to equal zero yields the Lyapunov dimension. Using a linear interpolation to find the 
precise value of this number is the commonly used Kaplan-Yorke formula~\cite{kaplan_yorke},
\begin{equation}
D_\lambda = J + \frac{S_J}{| \lambda_{J+1} |}
\end{equation}
where $J$ is the largest $j$ for which 
\begin{equation}
S_J=\sum_{j=1}^{J}\lambda_j > 0.
\end{equation}
The Lyapunov dimension $D_\lambda$  is an approximation to the number of degrees 
of freedom that contribute to the chaotic dynamics~\cite{farmer1983,cross:1993}.

The Karhunen-Lo\`{e}ve dimension was compared with the Lyapunov dimension 
by Sirovich and Deane~\cite{sirovich_deane} as the Rayleigh number was varied 
for turbulent Rayleigh-B\'{e}nard convection in a small periodic domain with free 
surface boundary conditions.  Zoldi~{\em et al.}~\cite{zoldi1998} computed $D_{KL}$ from experimental 
shadowgraph images of the spiral defect chaos state~\cite{morris:1993} of Rayleigh-B\'{e}nard 
convection in a large cylindrical domain with $\Gamma=109$. In order to explore 
the variation of $D_{KL}$ with the system size the data was spatially sampled 
with a window of increasing size.  Using this approach it was found that $D_{KL}$ 
scaled linearly with window size demonstrating extensivity.  For Rayleigh-B\'{e}nard 
convection in a shallow layer there are effectively two spatially extended directions and, 
as a result, the system size is measured as $\Gamma^2$.

The spiral defect chaos state has also been explored numerically.  Using a range of 
large periodic box geometries $D_\lambda$ was found to scale extensively with 
system size for $\Gamma=48,56,64$ and $R/R_c=1.8$~\cite{egolf:2000}. For box 
geometries the aspect ratio is defined as $\Gamma=L/d$ where $L$ is the length of 
an entire side of the box. The dimension $D_\lambda \approx 80$ for the largest domain 
and the dimension density was found to be $\delta_\lambda = D_\lambda/\Gamma^2 \approx 0.019$. 
The extensivity of chaos was also shown numerically for a range of finite cylindrical 
geometries for $4.72 \leq \Gamma \leq 15$, $\sigma=1$, and $R/R_c=3.5$~\cite{paul2007}. 
In this case the dimension density was found to be $\delta_\lambda \approx 0.25$.  
In our study we have chosen to explore the same system parameters as those of Ref.~\cite{paul2007} 
and to compute the variation of $D_{KL}$ over the same range of system sizes.
\begin{figure}[t!]
\begin{center}
    \includegraphics[height=2.5in]{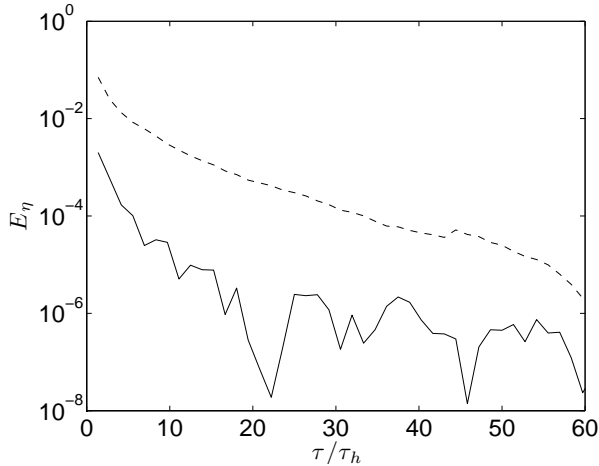}
\end{center}
\caption{The convergence of the normalized eigenvalue spectrum as a function of 
time $\tau$ for $\Gamma=6$ where the time has been normalized by the horizontal 
diffusion time $\tau_h$.  The error is given by $E_\eta$ for the method of 
snapshots (dashed line) and the Fourier method (solid line).}
\label{converge}
\end{figure}
\begin{figure}
\begin{center}
   \includegraphics[height=2.5in]{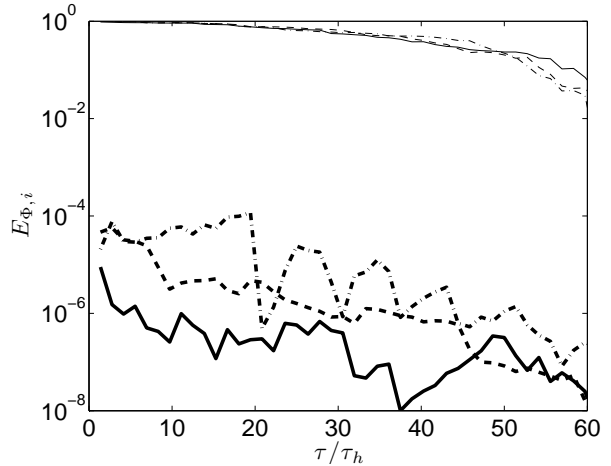}
\end{center}
\caption{The convergence of the eigenfunctions with time for a convection domain with 
aspect ratio $\Gamma=6$.  The error in the eigenfunction for mode $i$ is given by 
$E_{\Phi,i}$. Results are shown for $i=1$ (solid lines), $i=25$ (dashed lines), 
and $i=50$ (dash-dotted lines).  The results computed using the method of 
snapshots are represented by thin lines and those using the Fourier method 
are represented by thick lines.}
\label{mode_conv}
\end{figure}
  
\subsection{Convergence of the eigenvalues and eigenfunctions}
We next examine the convergence of the normalized eigenvalue distribution
\begin{equation}
\eta_i = \frac{\mu_i}{\displaystyle\sum_{i} \mu_i},
\end{equation}
where the index $i$ is over all computed eigenvalues $\mu_i$.  The variation of magnitude of the error with 
time $E_\eta(\tau)$ is computed as,
\begin{equation}
E_\eta(\tau) = 1 - \frac{|\eta_i(\tau)|}{|\eta_{i,\infty}|}
\end{equation}
where $|\eta_i|$ is the magnitude of the eigenvalue spectrum and $|\eta_{i,\infty}|$ is the value 
using the largest $\tau$ available for each case. The error $E_\eta$ decreases rapidly with time and is 
shown in Fig.~\ref{converge}.  The time has been normalized by the nondimensional time required for 
heat to diffuse a distance $r_0$ which we refer to as the horizontal diffusion time $\tau_h = \Gamma^2$. 
The nondimensional time scale of the fluid motion in a convection roll is on the order of $\tau \approx 1$ 
and $\tau_h$ represents a long-time scale. The convergence rate of the Fourier method is again found 
to be faster than the method of snapshots.  For  $\tau \approx 5 \tau_h$ the error is $E_\eta \sim 10^{-2}$ 
using the method of snapshots and $E_\eta \sim10^{-4}$ using the Fourier method. 
\begin{figure}[t!]
\begin{center}
    \includegraphics[height=2.5in]{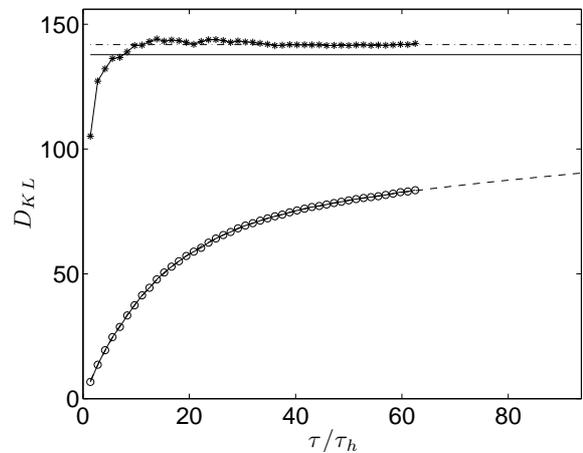}
\end{center}
\caption{The variation of $D_{KL}$ with time for the Fourier method ($\star$) and the method of snapshots 
($\circ$) for $\Gamma=6$.  The converged value for the Fourier method is given by the dash-dotted line. 
The curve fit for the method of snapshots is $D_{KL} = 137.8 -68.4 \exp{(-0.0730 \tau/\tau_h)} - 69.3 \exp{(-0.00406 \tau/\tau_h)}$ 
and is given by the dashed line. The asymptotic value of the dimension for the method of snapshots 
is given by the solid line at $D_\infty = 137.8$.}
  \label{KLvsSnap_g6}
\end{figure}

The convergence of the POD modes depend upon their eigenvalue and Fig.~\ref{mode_conv} 
shows the convergence of modes 1, 25, and 50.  Again we use the data for the largest available $\tau$ 
as an approximation for $\tau \rightarrow \infty$. The time variation of the magnitude of the error for the 
$i$th mode is computed as 
\begin{equation}
E_{\Phi,i}(\tau) = 1-\frac{(\bm{\Phi}_i(\tau),\bm{\Phi}_{i,\infty})}{(\bm{\Phi}_{i,\infty},\bm{\Phi}_{i,\infty})}.
\end{equation}
Again, the Fourier method converges faster than the method of snapshots.

\subsection{The convergence of the Karhunen-Lo\`{e}ve dimension}
The convergence of $D_{KL}$ for $\Gamma=6$ with respect to time is shown in Fig.~\ref{KLvsSnap_g6}. 
where the time has been scaled by the horizontal diffusion time $\tau_h$.  Results are given for both 
the Fourier method and the method of snapshots.  The Fourier method has reached a value of 
$D_{KL} \approx 142$ after $\tau \approx 20\tau_h$.

\begin{figure}[t!]
\begin{center}
    \includegraphics[height=2.5in]{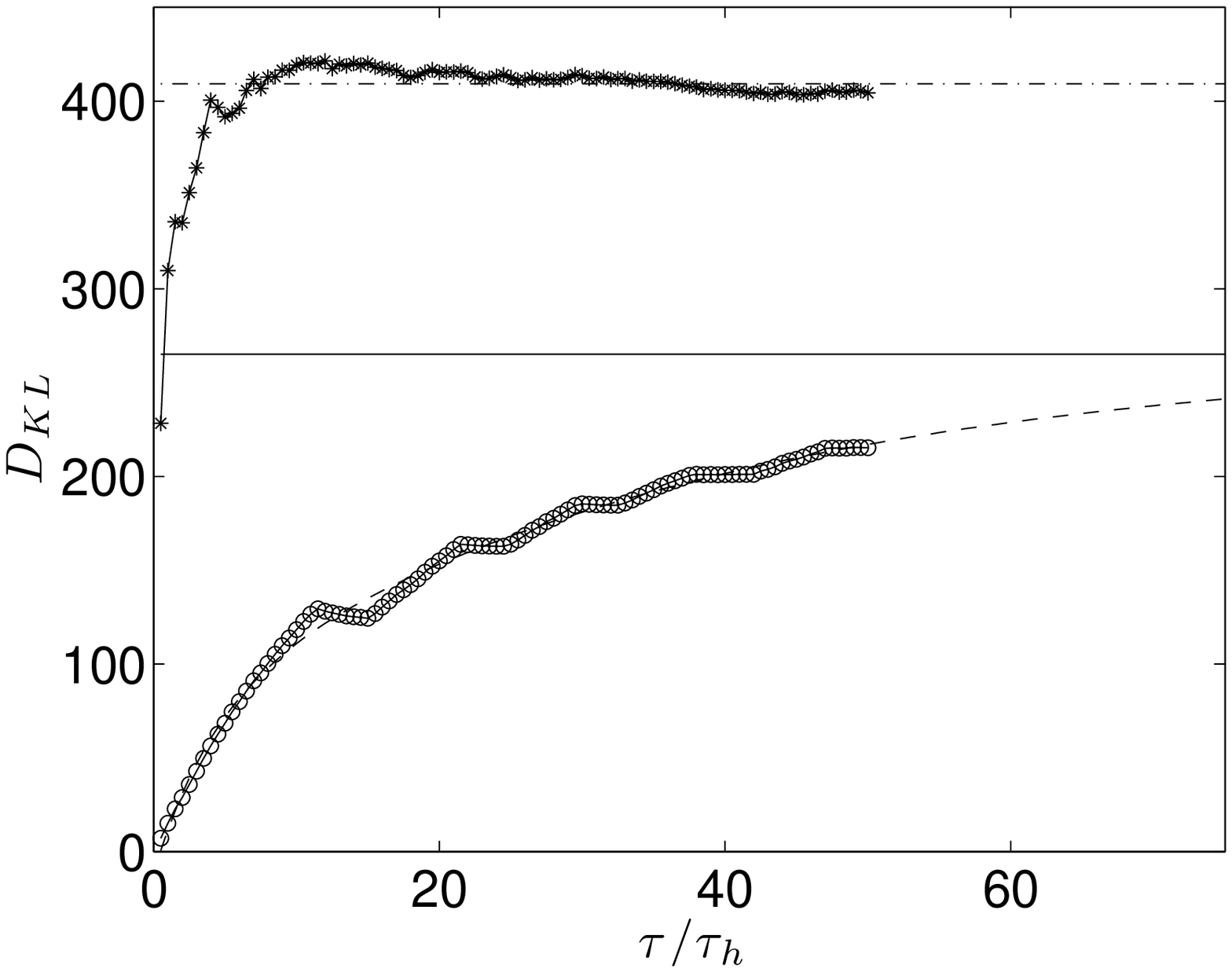}
\end{center}
\caption{The variation of $D_{KL}$ with time $\tau$ for the Fourier method ($\star$) and the 
method of snapshots ($\circ$) for $\Gamma=10$.  The converged value for the Fourier 
method is given by the dash-dotted line. The curve fit for the method of snapshots is  
$D_{KL}(\tau) = 265.2-82.0 \exp{(-0.2455 \tau/\tau_h)} - 194.8\exp{(-0.02796 \tau/\tau_h)}$ and is given 
by the dashed line. The asymptotic value of the dimension for the method of snapshots is 
given by the solid line at $D_\infty = 265.2$.}
\label{KLvsSnap_g10}
\end{figure}

\begin{figure}
\begin{center}
    \includegraphics[height=2.5in]{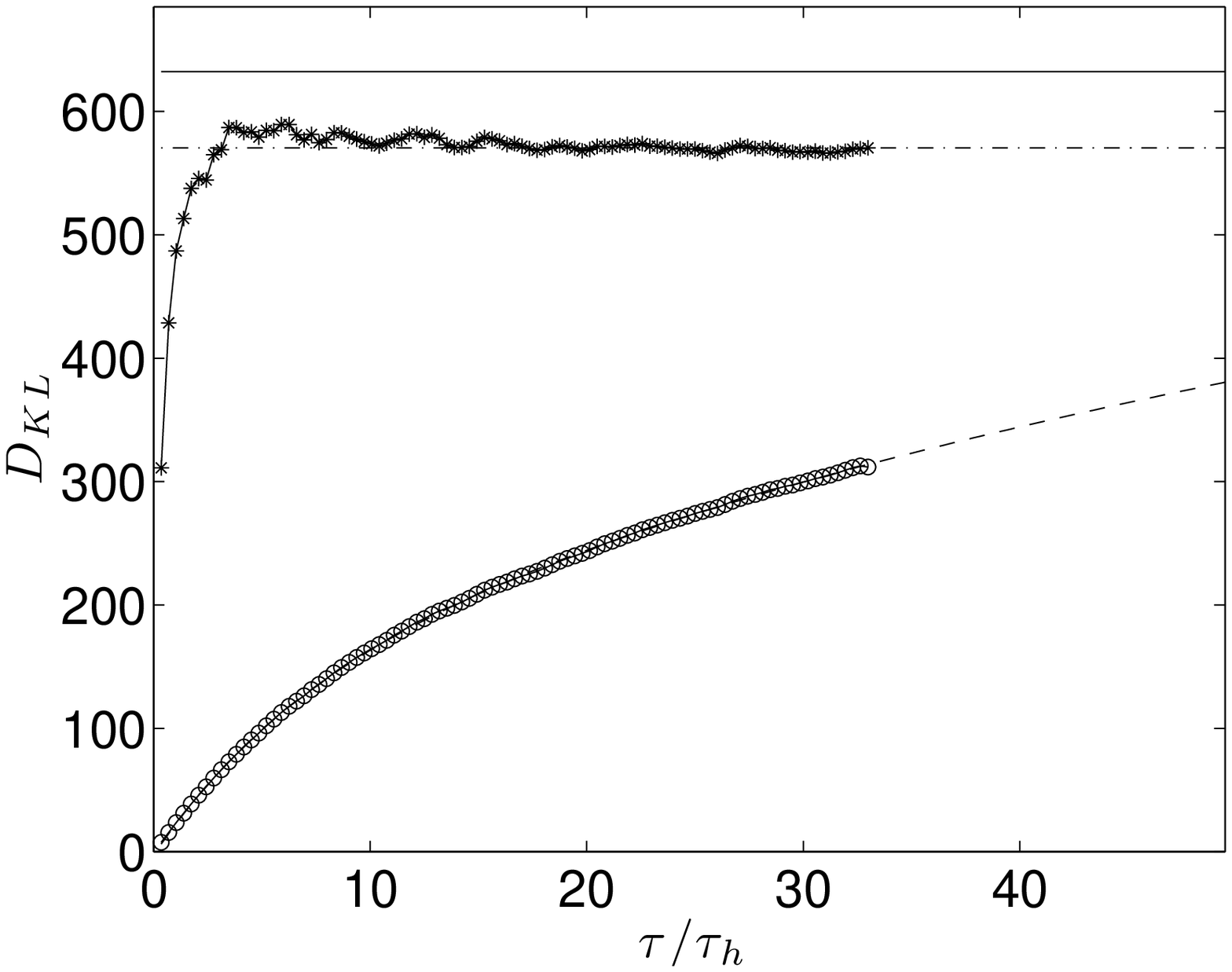}
\end{center}
\caption{The variation of $D_{KL}$ with time $\tau$ for the Fourier method ($\star$) and 
the method of snapshots ($\circ$) for $\Gamma=12$.  The converged value for the 
Fourier method is given by the dash-dotted line. The curve fit for the method 
of snapshots is $D_{KL}(\tau) = 632.2 -130.6\exp{(-0.1444 \tau/\tau_h)} - 504.2\exp{(-0.01404 \tau/\tau_h)}$ 
and is given by the dashed line. The asymptotic value of the dimension for the method 
of snapshots is given by the solid line at $D_\infty = 632.2$.  }
\label{KLvsSnap_g12}
  \end{figure}

\begin{figure}[t!]
  \begin{center}
    \includegraphics[height=2.5in]{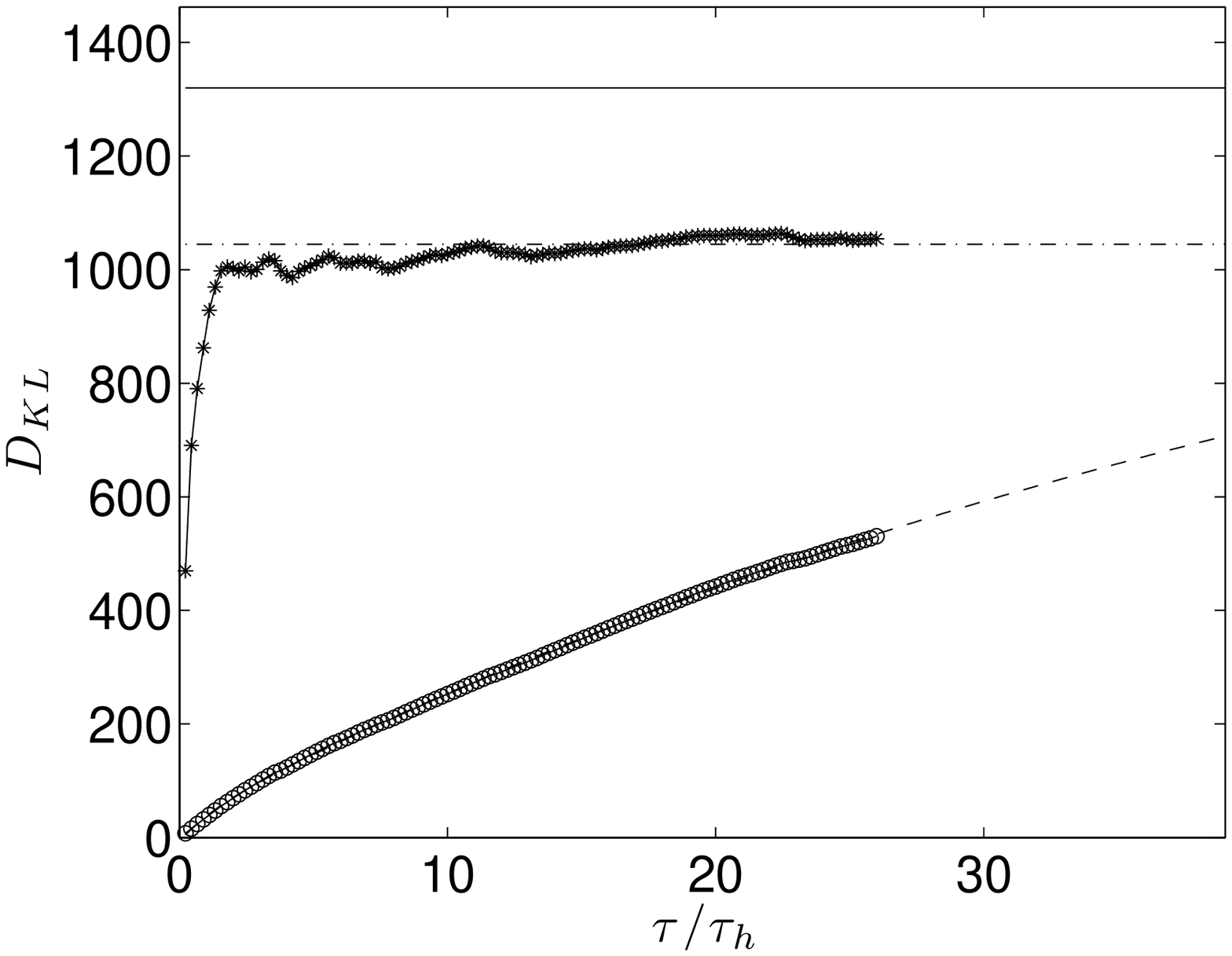}
\end{center}
\caption{The variation of $D_{KL}$ with time $\tau$ for the Fourier method ($\star$) and 
the method of snapshots ($\circ$) for $\Gamma=15$.  The converged value for the 
Fourier method is given by the dash-dotted line. The curve fit for the method 
of snapshots is $D_{KL}(\tau) = 1320.1-36.5\exp{(-0.7892 \tau/\tau_h)} - 1290.3\exp{(-0.01908 \tau/\tau_h)}$  
and is given by the dashed line. The asymptotic value of the dimension for the method of 
snapshots is given by the solid line at $D_\infty = 1320.1$.}
\label{KLvsSnap_g15}
\end{figure}

On the other hand, the method of snapshots exhibits a very gradual convergence.  The 
numerical simulations were continued until $\tau \approx 60 \tau_h$. In 
order to estimate a value of the asymptotic value of the dimension $D_\infty$ for the method of 
snapshots the results are fit with the following exponential dependence, 
\begin{equation}
D_{KL}(\tau) = D_\infty + c_1 \exp{(-c_2 \frac{\tau}{\tau_h})} + c_3 \exp{(-c_4 \frac{\tau}{\tau_h})}
\label{D_inf}
\end{equation}
which is shown as the dashed line.  The fitted curve is used to determine the asymptotic value 
of the dimension to yield $D_\infty =137.8$ and is indicated by the solid line. The 
convergence is very slow.  For the dimension to converge within 10\% of $D_\infty$ requires a 
time of $\tau \sim 400\tau_h$.  Such a slow convergence rate in time is typical of our results 
for the method of snapshots. A further difficulty resulting from the slow convergence is the that 
the value of $D_\infty$ is very sensitive to small variations in the data that is made more significant 
when only a finite amount of data is available.  The variation of $D_{KL}$ with time is shown for 
$\Gamma=10, 12,15$ in Figs.~\ref{KLvsSnap_g10}-\ref{KLvsSnap_g15} respectively and numerical 
values of the $D_{KL}$ are given in Table~\ref{dimensions}. 
\begin{table}[t]
\caption{The variation of the dimension with system size.  The asymptotic values of the Karhunen-Lo\`{e}ve 
dimension $D_\infty$ are given for the method of snapshots and the Fourier method. The Lyapunov 
dimension $D_\lambda$ computed for the same system parameters are included from Ref.~\cite{paul2007}.}
\begin{center}
\begin{tabular}{llll}
\hline
                   & Snapshots & Fourier Method & Lyapunov \\
$\Gamma$ & $D_{KL}$  & $D_{KL}$           & $D_\lambda$  \\
\hline
6 & 138     & 142  & 6.8 \\
10 & 265   & 409 & 21.6 \\
12 & 632   & 570 & 32.5  \\
15 & 1320 & 1045 & 53.7 \\
\hline
\end{tabular}
\end{center}
\label{dimensions}
\end{table}
\subsection{Extensivity}
\begin{figure}
\begin{center}
    \includegraphics[height=2.5in]{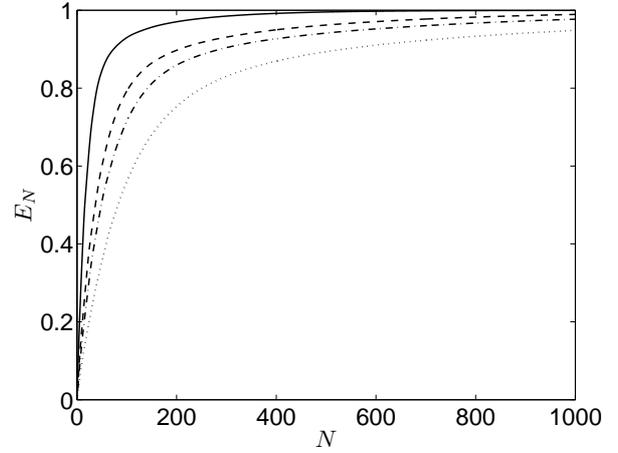}
  \end{center}
\caption{The fraction of the total energy captured $E_N$ as a function of the number 
of modes $N$ using the Fourier method. Results are shown for $\Gamma=6$ (solid line), 
$\Gamma=10$ (dashed line), $\Gamma=12$ (dash-dotted line), and $\Gamma=15$ (dotted 
line).  A typical approach used to compute the value of $D_{KL}$ is to use the number of 
modes needed for $E_N=0.9$. From these results it is clear that $D_{KL}$ increases in 
value as the system size is increased.}
 \label{running_total}
\end{figure}
\begin{figure}[t!]
\begin{center}
   \includegraphics[height=2.5in]{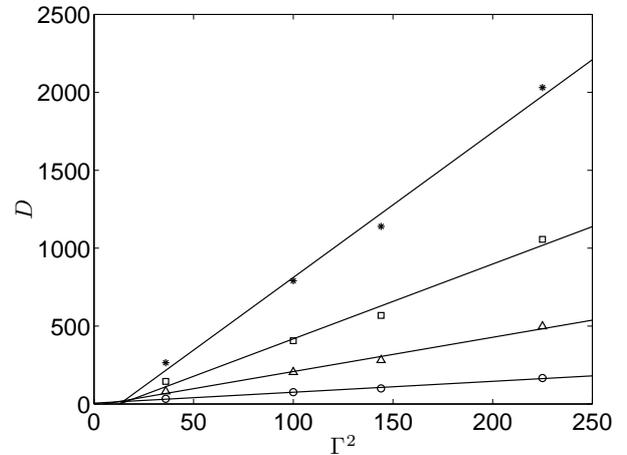}
\end{center}
\caption{The variation of the dimension $D_{KL}$ with system size $\Gamma^2$ for several choices 
of $f$ using the Fourier method. Results are shown for $f=0.5$ (circles), $f=0.8$ (triangles), 
$f=0.9$ (squares), and $f=0.95$ (stars). The solid lines are linear curve fits through the 
data indicating the extensivity of $D_{KL}$. In the following figures we have chosen to 
use $f=0.9$ when reporting values of $D_{KL}$.}
  \label{variance}
 \end{figure} 
\begin{figure}
\begin{center}
    \includegraphics[height=2.5in]{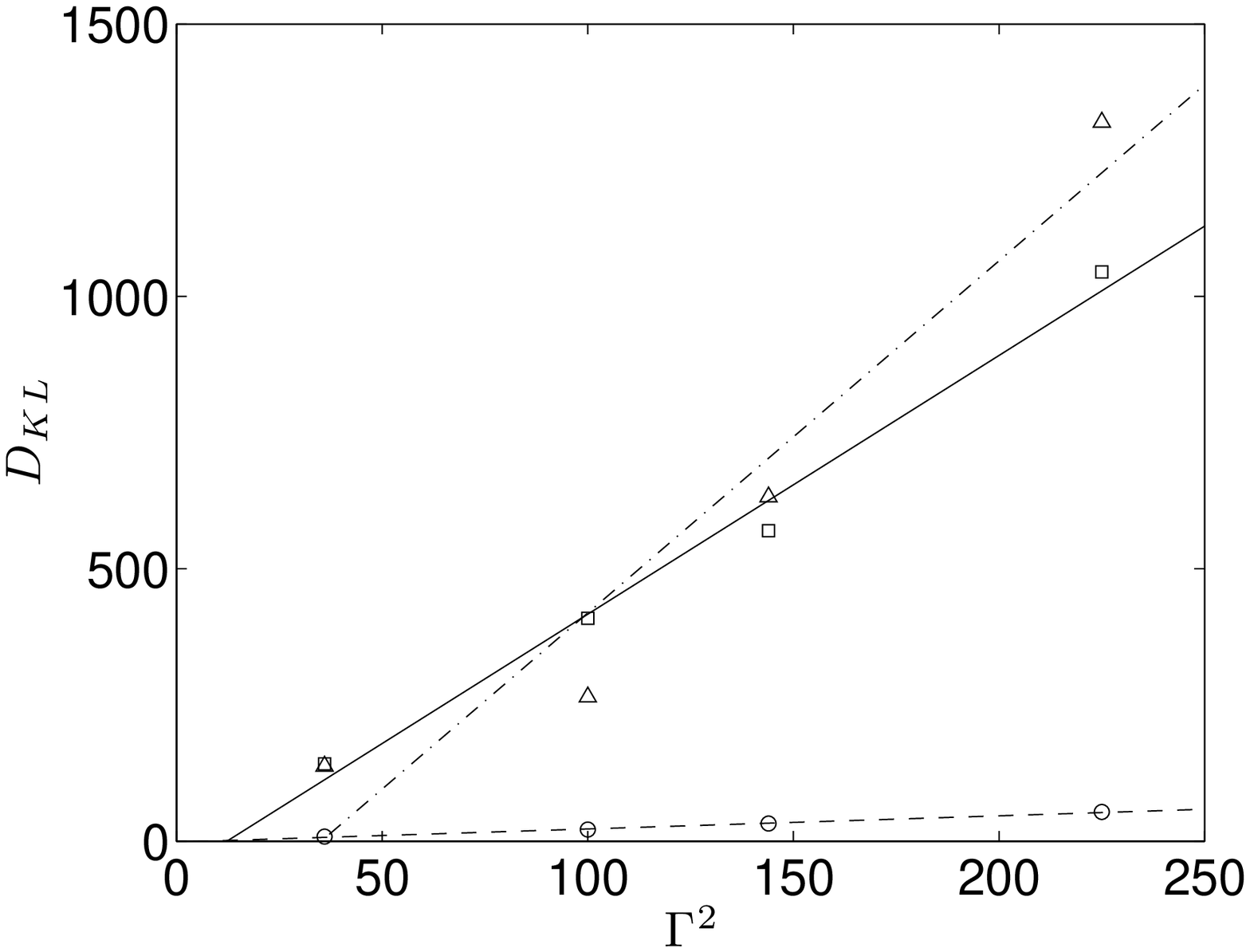}
\end{center}
\caption{The variation of $D_{KL}$ and $D_\lambda$ with system size $\Gamma^2$ using numerical 
results for $\Gamma=6, 10, 12$ and $15$.  The values of $D_{KL}$ computed using the Fourier method 
are represented as squares and the solid line is a linear curve fit. The value of $D_{KL}$ computed 
using the method of snapshots are represented as triangles and the dash-dot line is a linear 
curve fit. The Lyapunov dimension from Ref.~\cite{paul2007} are represented as circles and the 
dashed line is a linear curve fit. From the data $D_{KL} \approx 19.7 D_\lambda$ where the Fourier 
method has been used for $D_{KL}$.}
  \label{KLvsLyap}
\end{figure}

Figure~\ref{running_total} shows the time variation of the cumulative normalized energy of the first $N$ 
eigenvalues.  The normalized energy is computed from,
\begin{equation}
E_N(\tau) = \frac{\displaystyle\sum_{k=0}^{N} \mu_k } {\displaystyle\sum_{k=0}^{\infty} \mu_k}
\end{equation} 
using the data from the Fourier method with the largest $\tau$ for each respective system size.  The basic trend of 
$E_N(\tau)$ shows that the total number of modes $N$ needed to capture a fraction $f$ 
increases with system size.  Figure~\ref{variance} shows the variation of $D_{KL}$ with system 
size using different values of $f$ where $0.5 \le f \le 0.95$.  Our results yield a linear relationship 
between $D_{KL}$ and $\Gamma^2$ over the entire range. The choice for $D_{KL}$ that we 
have used for reporting values of $D_{KL}$ is $f=0.9$.  

The variation of $D_{KL}$ with system size is shown in Fig.~\ref{KLvsLyap} using the 
Fourier method (squares) and the method of snapshots (triangles) and linear curve fits 
through the data are given by the solid line and the dash-dotted line, respectively.  The results 
using the method of snapshots increase with system size but with with significant deviations 
from the line of extensivity.  As discussed, the rate of convergence for the method of snapshots 
is very slow and our numerical results would need to be continued for much longer simulation 
times for these results to converge. However, if this were done the results would eventually 
agree with those of the Fourier method which is extensive. 

The variation of $D_\lambda$ with system size is shown in Fig.~\ref{KLvsLyap} by the circles 
with a linear curve fit given by the dashed line. Although there is no reason to expect the two 
dimensions to be related since they measure very different aspects of the patterns and their 
dynamics, it has been suggested that for extensively chaotic dynamics their rate of 
increase should be similar~\cite{zoldi1997}. Numerical values of the dimensions are 
given in Table~\ref{dimensions}.  Using the linear curve fits through the data yields 
the relationship $D_{KL} \approx 19.7 D_{\lambda}$.

\section{Conclusions}
We have performed a proper orthogonal decomposition on a finite cylindrical domain 
containing a shallow fluid layer undergoing extensively chaotic dynamics.  Our results suggest that 
even for this case the required time necessary to obtain good estimates of the Karhunen-Lo\`{e}ve dimension 
are very long and on the order of hundreds of horizontal diffusion times with the commonly used 
method of snapshots. However, by exploiting the rotational symmetry of the problem the time to 
convergence can be drastically reduced using the Fourier method. This has important consequences 
for more complex flow fields that are common in engineering applications where the amount of 
data available is limited in both experiment and for computations.  A prime example is the use of 
particle image velocimetry to obtained detailed information about the flow field. In this case the observation 
time is a function of the amount of data that can be captured by the camera.  Overall, our results 
show that one must be careful to ensure the convergence of the proper orthogonal decomposition 
in order to obtain an accurate estimation of the dimension.

\section*{Acknowledgments}
This computations were conducted using the resources of the National Science Foundation 
TeraGrid and the Advanced Research Computing center at Virginia Tech.  MRP acknowledges support 
from NSF grant no. CBET-0747727. We also gratefully acknowledge many useful interactions 
with Paul Fischer and Mike Cross.  

\bibliographystyle{elsarticle-num}

\bibliography{PODtime.bbl}

\end{document}